\documentclass{aa}  

\usepackage{graphicx}
\usepackage{txfonts}
\usepackage{xcolor}
\usepackage{hyperref}

\def\Ha{H$\alpha$}

\def\kms{km~s$^{-1}$}
\newcommand{\hi}{H\,{\sc i}}
\newcommand{\hii}{H\,{\sc ii}}

\newcommand{\OIII}{\mbox{[O\,\textsc{iii}]}}

\newcommand{\NII}{\mbox{[N\,\textsc{ii}]}}
\newcommand{\SII}{\mbox{[S\,\textsc{ii}]}}

\def\dis        {{\em DIS}\/}

\def\bino       {{\em Binospec}\/}
\def\apo        {{\em APO}\/}

\def\ga         {{NGC~4569}\/}
\def\gb         {{NGC~4531}\/}

\begin{document} 

   \title{A Virgo Environmental Survey Tracing Ionised Gas Emission (VESTIGE). XIX. The discovery of a spectacular 230 kpc \Ha{} tail following \ga{} in the Virgo cluster}
   \titlerunning{\ga's long H$\alpha$ tail}
   
   \author{M. Sun\inst{1},
          H. Le\inst{1},
          B. Epinat\inst{2,3},
          A. Boselli\inst{2,4},
          R. Luo\inst{5},
          K. Hosogi\inst{1},
          N. Pichette\inst{1},
          W. Forman\inst{6},
          C. Sarazin\inst{7},
          M. Fossati\inst{8,9},
          H. Chen\inst{10},
          G. Hensler\inst{11},
          E. Sarpa\inst{12},
          P. Amram\inst{2},
          J. Braine\inst{13},
          J. C. Cuillandre\inst{14},
          S. Gwyn\inst{15},
          S. Martocchia\inst{2},
          B. Vollmer\inst{16}
          }
   \authorrunning{M. Sun et al.}

   \institute{Department of Physics \& Astronomy, University of Alabama in Huntsville, 301 Sparkman Dr NW, Huntsville, AL 35899, USA\\
              \email{ms0071@uah.edu}
         \and
             Aix Marseille Universit\'{e}, CNRS, LAM (Laboratoire d’ Astrophysique de Marseille) UMR 7326, 13388, Marseille, France
         \and
             Canada-France-Hawaii Telescope, 65-1238 Mamalahoa Highway, Kamuela, HI 96743, USA
         \and
             INAF - Osservatorio Astronomico di Cagliari, Via della Scienza 5, 09047 Selargius (CA), Italy
         \and
            School of Physics and Electronic Science, Guizhou Normal University, Guiyang 550001, People's Republic of China; Guizhou Provincial Key Laboratory of Radio Astronomy and Data Processing, Guizhou Normal University, Guiyang 550001, People's Republic of China
         \and
             Center for Astrophysics | Harvard \& Smithsonian, 60 Garden St., Cambridge, MA 02138, USA
         \and
             Department of Astronomy and Virginia Institute for Theoretical Astronomy, University of Virginia, P.O. Box 400325, Charlottesville, VA 22904, USA
         \and
             Dipartimento di Fisica G. Occhialini, Universit\`{a} degli Studi di Milano Bicocca, Piazza della Scienza 3, I-20126 Milano, Italy
         \and
             INAF-Osservatorio Astronomico di Brera, via Brera 28, I-20121 Milano, Italy
         \and
             Research Center for Astronomical Computing, Zhejiang Laboratory, Hangzhou 311100, People’s Republic of China
        \and
             Department of Astrophysics, University of Vienna, T\"urkenschanzstrasse 17, 1180, Vienna, Austria
         \and
             SISSA – International School for Advanced Studies, Via Bonomea 265, 34136 Trieste, Italy
         \and
             Laboratoire d'Astrophysique de Bordeaux, Univ. Bordeaux, CNRS, B18N, all\'ee Geoffroy Saint-Hilaire, 33615, Pessac, France
         \and
             AIM, CEA, CNRS, Universit\'e Paris-Saclay, Universit\'e Paris Diderot, Sorbonne Paris Cit\'e, Observatoire de Paris, PSL University, F-91191 Gif-sur-Yvette Cedex, France
         \and
             National Research Council of Canada, Herzberg Astronomy and Astrophysics, 5071 West Saanich Road, Victoria, BC, V9E 2E7, Canada
         \and
             Universit\'e de Strasbourg, CNRS, Observatoire astronomique de Strasbourg, UMR 7550, 67000, Strasbourg, France
             }

  \abstract
   {Galaxies fly inside galaxy clusters and ram pressure by the intracluster medium (ICM) can remove a large amount of the interstellar medium (ISM) from the galaxy, and deposit the gas in the ICM. The ISM decoupled from the host galaxy leaves a long trail following the moving galaxy. Such long trails track the galaxy motion and can be detected with sensitive data in H$\alpha$.}
   {We study the H$\alpha$ tail trailing \ga{} in the Virgo cluster.}
   {The initial discovery was made with the deep H$\alpha$ imaging data with CFHT, from the VESTIGE project. The follow-up spectroscopic observations were made with \apo{}/{\em DIS}, {\em MMT}/\bino{} and {\em CFHT}/{\em SITELLE}.}
   {Besides the known 80 kpc H$\alpha$ tail downstream of \ga{}, the deep H$\alpha$ imaging data allow the H$\alpha$ tail detected to at least 230 kpc from the galaxy. More importantly, the H$\alpha$ clumps implied from the imaging data are confirmed with the spectroscopic data. The H$\alpha$ clumps show a smooth radial velocity gradient across $\sim$ 1300 km/s, eventually reaching the velocity of the cluster.} 
   {This discovery, for the first time, demonstrates the full deceleration process of the stripped ISM. This discovery also showcases the potential with wide-field H$\alpha$ survey on galaxy clusters to discover intracluster optical emission-line clouds originated from cluster galaxies. These clouds provide kinematic tracers to the infall history of cluster galaxies and the turbulence in the ICM. They are also excellent multi-phase objects to study the classical cloud crushing problem and other relevant important physical processes.
   }


   \maketitle

\begin{table*}
        \begin{center}
                \caption{The \bino{}/{\em MMT} observations}
                \label{tab:observational info.}
                \vspace{-0.3cm}
                \begin{tabular}{cccccccc}
                        \hline
                        Name & RA & Dec & PA & \# of slits (width) & Date & Exposure & Condition (seeing, airmass) \\
                             & (hh mm ss) & ($\degr$ $\arcmin$ $\arcsec$) & & & & (sec) & \\
                        \hline
                        field1 & 12 35 57.95 & +13 11 53.70 & 19 & 37+14 (1$''$), 22 & 03/07/2022 & 600$\times$4 & clear (1.1$''$-1.4$''$, 1.11) \\
                        field2 & 12 34 52.00 & +13 09 20.80 & -90 & 33+39 (1$''$), 36 & 03/25/2022 & 900$\times$7 & clear (0.8$''$, 1.06) \\
                        field3 & 12 35 46.78 & +13 03 08.73 & -83 & 23+14 (1.5$''$), 26 & 03/25/2023 & 900$\times$8 & clear (0.9$''$-1.7$''$, 1.06) \\
                        field4 & 12 33 51.81 & +12 58 57.73 & 89 & 10+24 (1.5$''$), 10 & 01/28/2023 & 900$\times$7 & thin clouds (1.5$''$-1.9$''$, 1.10) \\
                        \hline
                \end{tabular}
        \end{center}
        \vspace{-0.2cm}
        Note: Observations are taken under the program: SAO-12-22A, SAO-23-22B and SAO-25-23A (PI: Forman/Sun). The listed slit number is for side A + side B, with the last number as the number of robust detections used in this paper.
    \label{t:obs}
\end{table*}

\section{Introduction}

Most baryons in galaxy clusters are in the hot ($T \sim 10^{7} - 10^{8}$ K) intracluster medium (ICM) that permeates the space between galaxies. As cluster galaxies soar through the ICM, the interaction with the ICM plays a vital role in galaxy evolution, through ram pressure stripping (RPS) of the interstellar medium (ISM) \citep[e.g.,][] {Boselli22}.
The ISM, removed from the galaxy by RPS, forms tails downstream of the moving galaxy. Stripped tails have been detected and studied in CO, radio continuum, \hi{}, H$\alpha$ and X-rays (see the recent summary in \citealt{Boselli22}).
Tails vary in surface brightness, typically with bright clumps/filaments embedded in a more diffuse, uniform component. 
Star formation (SF) activity also varies in tails with the SF conditions in tails not well understood. 
So far, one of the best probes on stripped tails is H$\alpha$, with a large covering fraction, good sensitivity, and high angular resolution together \citep[e.g.,][]{Gavazzi01,Yagi10,Poggianti17}, while the data at some other bands either lack sensitivity and angular resolution (e.g., \hi{}, X-rays), or have a small covering fraction (e.g., CO).

Stripped tails can trace the infall history of cluster galaxies, especially if they can be detected far from the galaxy. On the other hand, stripped ISM is subject to the cluster ``weather'' so their studies can provide constraints on the ICM turbulence. Thus, long stripped tails are valuable laboratories. However, nearly all stripped tails have projected lengths below $\sim$ 100 kpc (see summary in \citealt{Boselli22}).
In this paper, we present the discovery of a 230 kpc long H$\alpha$ tail following \ga{} in the Virgo cluster, also with the kinematic confirmation to recover the deceleration history of the stripped ISM.
\ga{} is one of the largest and most massive spirals in the Virgo cluster.
It has a remarkably high velocity with respect to the Virgo cluster with a radial velocity of -235 km s$^{-1}$ \citep{1991rc3..book.....D}.
\cite{Kashibadze20} gave a velocity of 1070 km~s $^{-1}$ for the Virgo cluster, while the velocity of M87 is 1284 km/s \citep{Cappellari11}. 
Thus, \ga{}'s relative velocity along the line of sight is $\sim$ 1300 km/s, while its long H$\alpha$ and X-ray tails (\citealt{Boselli16,Sun22}; and this work) also shows a significant velocity component in the plane of sky.
\ga{} has lost over 90\% of its initial \hi{} gas \citep[e.g.,][]{Cayatte90,Vollmer04}.
It has a truncated disk observed in CO, HI, H$\alpha$ and dust and shows the impact of RPS for several hundred Myr \citep[e.g.,][]{Boselli06,Boselli16}.
An ionized gas tail has been revealed to $\sim$ 80 kpc (projected) from the galaxy \citep{Boselli16}, which motivated the Virgo Environmental Survey Tracing Ionised Gas Emission (VESTIGE) \citep{Boselli18}.
With the deeper VESTIGE data, this paper presents the discovery of an even longer H$\alpha$ tail trailing \ga{}. More importantly, the follow-up optical spectroscopic data confirm the association of many clumps/filaments in the tail with \ga{} and also show a coherent velocity gradient as these stripped clumps are decelerated by ram pressure.
This paper presents the initial discovery and the velocity confirmation.
More detailed studies on the ionized gas clumps (e.g., line ratios, local kinematics, velocity dispersion, velocity structure function) with the \bino{} and {\em MUSE} data will be presented in a future paper.
We adopted a distance of 16.5 Mpc to the Virgo cluster \citep{Mei07}. Therefore, 1$'$ = 4.80 kpc.

\begin{figure*}[t!]
\vspace{-0.3cm}
\hspace{-1.9cm}
\includegraphics[width=1.11\textwidth]{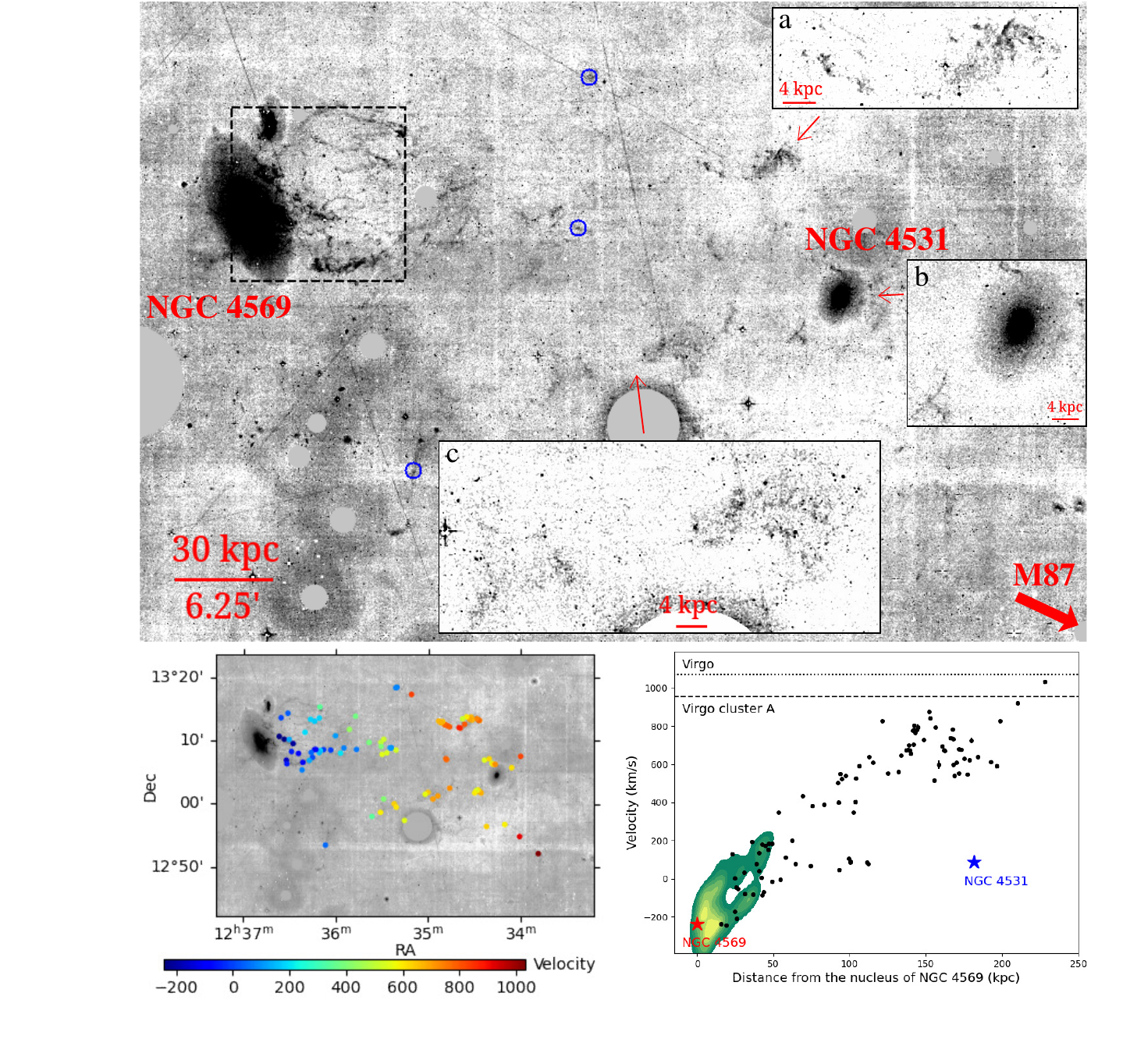}
\vspace{-1.7cm}
\caption{
{\em Upper:}
The net H$\alpha$ image showing the extended optical emission-line clumps/filaments trailing \ga{}. Bright stars and their scattered halos are masked. 
The black box in the dashed line is the {\em SITELLE} field shown in Fig.~\ref{fig:sitelle}. Three zoom-ins are also shown. 
The direction towards M87 is also shown at the lower right corner.
Note that the clumps/filaments around NGC~4531 have very different velocities from NGC~4531. Because of the various instrumental artifacts, spectroscopic follow-ups are required.
{\em Lower Left:}
The same net H$\alpha$ image as the one shown in the upper panel, with velocities of 94 clumps from \bino{} overlaid.
See Fig.~\ref{fig:sitelle} for velocities at the front part of the tail from {\em SITELLE}. 
{\em Lower Right:}
The velocities (and 1-$\sigma$ uncertainty) of 94 clumps from \bino{} in black + the kernel density estimation (KDE) of the {\em SITELLE} velocities (as shown in Fig.~\ref{fig:sitelle}) vs. distance from the nucleus of \ga{}.
The KDE shows the probability density function of the {\em SITELLE} velocities, with yellow as the highest density.
For reference, \ga{} is at -235~\kms and \gb{} is at 90~\kms. The velocity of the Virgo cluster A is 955~\kms~ from \cite{Boselli14}, while \cite{Kashibadze20} gave a velocity of 1070~\kms~ for the Virgo cluster.
The observed clumps show a smooth velocity gradient from \ga{} and there is no evidence for their connection with \gb{}. The end of the detected H$\alpha$ tail is nearly at rest (along the line of sight) with respect to the Virgo cluster mean velocity.
There are a few clumps at distance $\sim$ 100 kpc with velocities of 50 - 100~\kms. They are in three groups and are marked in small blue circles in the top panel. Despite their similar velocities to \gb{}'s, there is no spatial evidence for the connection.
}
\label{fig:f1}
\end{figure*}

\begin{figure}
\vspace{-0.2cm}
\hspace{-0.55cm}
\includegraphics[width=0.57\textwidth]{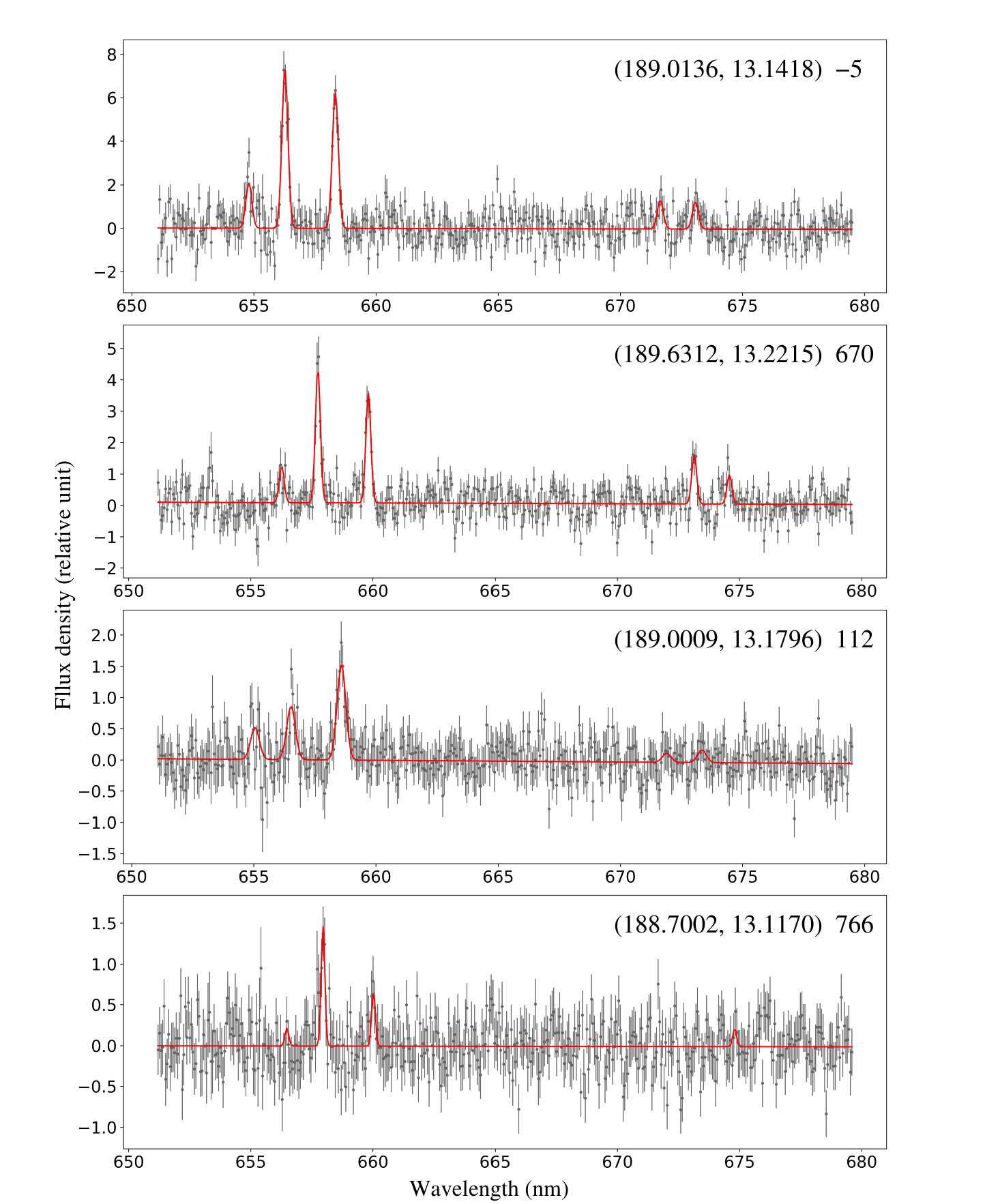}
\vspace{-0.5cm}
\caption{
\bino{} spectra on four slit positions, from one of the brightest regions in the top, a region with median brightness in the 2nd row, to two spectra with about the least significant detections in our sample. Our best-fit model is shown in the red line. Only the portion around \NII{}, \Ha{} and \SII{} is shown. The shown spectra are unbinned on the spectral axis with a 0.62\AA{} sampling.
The central coordinate of each position (in brackets) and the best-fit velocity are shown in the upper right corner.
}
    \label{fig:spec}
\end{figure}

\begin{figure}
\hspace{-0.25cm}
\includegraphics[width=0.5\textwidth]{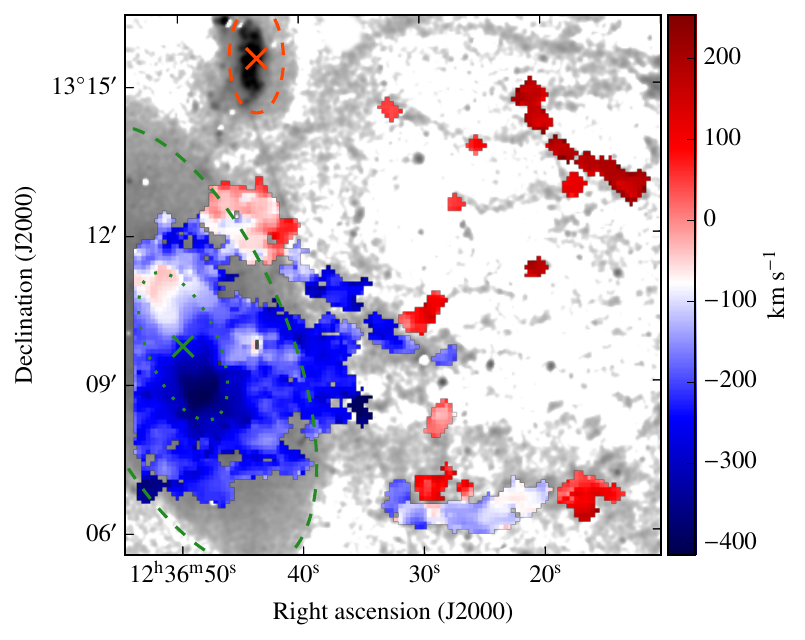}
\vspace{-0.6cm}
\caption{
The {\em SITELLE} velocity map in the region shown in Fig.~\ref{fig:f1}, overlaid on the net H$\alpha$ image in the grey scale.
The big green ellipse in the dashed line shows the D25 aperture of \ga{}, while the small green ellipse in the dotted line is 1/3 of the size of the big green ellipse. \ga{}'s nucleus is marked by a green cross. One can observe the disk rotation close to the nucleus and the truncated H$\alpha$ disk \citep[e.g.,][]{Boselli16}. The red ellipse in the dashed line shows the D25 aperture of IC~3583 with a velocity of 1121 km s$^{-1}$. Its interaction with \ga{} was excluded by \cite{Boselli16}.
}
    \label{fig:sitelle}
\end{figure}

\section{VESTIGE data and the long H$\alpha$ tail trailing \ga{}}

\cite{Boselli16} presented results based on 1.28 hours of narrow-band imaging data with {\em CFHT}/{\em MegaCam}, centered around the near tail region of \ga{}.
VESTIGE adds 2 hours of narrow-band imaging data covering a much wider field around \ga{} \citep{Boselli18}.
The deeper H$\alpha$+\NII{} data and the wider coverage allow the detection of the longer extension to over $\sim$ 200 kpc from \ga{} (Fig.~\ref{fig:f1}). The H$\alpha$ tail is composed of many filaments and clumps, but lacking bright, compact regions like \hii{} regions. The average distance between clumps/filaments also increases beyond the initial 80 kpc tail. There is no detected diffuse H$\alpha$ component with a high covering fraction to
connect clumps/filaments. However, such a diffuse, low surface brightness tail would be difficult to be robustly detected with all the image artifacts, e.g., scattered light around bright stars, increased noise along the chip gaps, and the flat field uncertainty at large scales.
Because of all the artifacts (evident in Fig.~\ref{fig:f1}), spectroscopic observations are required to confirm the existence of these emission-line features and their association with \ga{}, as discussed
in the following sections.

\section{\bino{} data and analysis}

We conducted spectroscopic observations of emission-line objects trailing \ga{} with the \bino{} spectrograph on the {\em MMT} telescope (Table 1). 
\bino{} is a multislit imaging spectrograph with wide wavelength coverage and high efficiency \citep{Bino19}. It has two sides, each with a field of view of 8$' \times 15'$ separated by a gap of 3.2$'$. Each side is covered by a sheet metal slit mask that can hold over 150 slitlets, parallel to the long side of the mask. We used the 600 lines/mm grating. The wavelength coverage varies across the field but is typically  $\sim$ 4900 \AA - 7200 \AA, with a 0.62 \AA/pixel dispersion and $\sim$ 2800 resolving power for 1$''$ slits.
The \bino{} data were reduced with the \bino{} pipeline\footnote{\url{https://bitbucket.org/chil_sai/binospec/src/master/}}.
As our targets are diffuse clouds that can fill individual slitlets, we did not use the automatic spectrum extraction in the pipeline. Instead, the 2D spectrum of each slit was examined to identify the regions of interest for extraction, also with the help of the VESTIGE imaging data. The spatial dimension of the spectral extraction regions ranges from 2.9$''$ to 18.2$''$, with a median length of 6.7$''$ (or 0.54 kpc).

We then used our own python code to extract 1D spectra from the 2D spectra and fit spectra to obtain barycentric velocities and line ratios. The fit was done with five Gaussian functions with the same velocity to model the H$\alpha$, \NII{} and \SII{} lines, plus a continuum model.
\NII{} $\lambda$ 6584 / \NII{} $\lambda$ 6548 is fixed at 3.
While H$\beta$ and \OIII{} wavelengths are covered by about half of the data,
there are only a few robust detections.
LMFIT\footnote{\url{https://lmfit.github.io/lmfit-py/}}
was used for the non-linear least-squares minimization and the fitting methods are dual annealing and Levenberg-Marquardt.
Examples of four spectra are shown in Fig.~\ref{fig:spec}.
We also verified our velocity results by running RVSAO on a small subsample and found consistent velocity measurements. 
In total, we obtained robust velocities for 94 emission-line objects, all with velocities consistent with the objects in the Virgo cluster (Fig.~\ref{fig:f1}).
The median velocity statistical error is 5 km s$^{-1}$, while the velocity systematic uncertainty is 6 - 10 km s$^{-1}$.
There are 137 slits in total for candidate H$\alpha$ clouds.
The success rate, defined as the fraction of slits with derived velocities consistent with that of the Virgo cluster, is $\sim$ 2/3.
Seventeen slits have weak line signals but below our detection threshold ($> 3 \sigma$ for at least two lines). Excluding the shallow field1 data, the success rate is $\sim$ 80\%, demonstrating the robustness of the VESTIGE H$\alpha$ imaging data, even near the detection threshold.
We have also obtained {\em VLT/MUSE} data in three fields at 60, 171 and 210 kpc from the nucleus (program: 111.259P; PI: Sun). There are five velocities measured from \bino{} in these three {\em MUSE} fields and those velocity values are confirmed by the {\em MUSE} data, as well as the line ratios. Results from the {\em MUSE} data will be presented in a future paper.
The MW extinction has been corrected with the \cite{Fitzpatrick99} extinction law, with $A_{\rm V} = 0.13$ mag and $R_{\rm V}$ = 3.1.

\section{\apo{} data and analysis}

While most the spectroscopic results in this work come from \bino{}, the spectroscopic follow-up started with the Dual Imaging Spectrograph (\dis{}) on the Apache Point Observatory (\apo{}) in 2021 (PI: Sarazin/Sun). On Jan. 16 and Feb. 4, 2021, we observed a cloud at (RA, DEC) = (12:34:31.7, +13:13:46) for 6$\times$10 min with a 2$''$ slit and 12$\times$10 min with a 5$''$ slit respectively.
\NII{}, \SII{} and H$\alpha$ lines at a velocity of $\sim$ 650~\kms\, were clearly detected.
As this region was later well covered by \bino{} with $\sim 3\times$ depth, we do not include the \dis{} results in the velocity analysis. The \dis{} spectra were used to independently verify the \bino{} velocities in the same region, as well as the line ratios.
More importantly, this initial confirmation helped to justify the later \bino{} observations.  

\section{{\em SITELLE} data and analysis}

{\em SITELLE} is an imaging Fourier Transform Spectrometer (iFTS) that produces spectra for any 0.32$''\times0.32''$ pixels within its 11$'\times11'$ field of view \citep{Drissen19}. It was used on February 9-11, 2024 to observe \ga{} and its tail with a 12.3 nm wide (at 90\% transmission) filter SN4 specifically designed to allow the detection of diffuse \Ha{} emission down to $5\times 10^{-18}$ erg~s$^{-1}$~cm$^{-2}$~arcsec$^{-2}$ after a 16$\times 16$ pixels (5\arcsec $\times$ 5\arcsec) binning in about 5 hours of on-source exposures (commissioning data). 
The field is shown in Fig.~\ref{fig:f1}. The spectral resolution was set to R=4000, using 375 scanning steps of 50 sec on-source exposure each. The sky conditions were clear with 0.8$'' - 1.5''$ seeing.

The data was reduced using the standard ORB/ORBS software used at CFHT \citep{Martin15}. A dedicated analysis was performed to remove sky lines prior to further analysis. Median sky spectra computed over extended areas were removed for each pixels. The velocity variations across the tails ensures that the associated emission lines are not present in the median sky spectra. The data has also been calibrated in wavelength using the OH night sky lines to provide accurate barycentric velocity measurements; to reach a $\sim 10-40$~\kms{} accuracy depending on the line intensities.

In order to enhance the signal on faint extended regions, the data cube was smoothed using a 32-pixels FWHM Gaussian kernel and binned by a factor 16. It was also spectrally smoothed with a 2-pixels FWHM Gaussian kernel before analysis. The emission line analysis was performed using a version of the \emph{Camel} software\footnote{\url{https://gitlab.lam.fr/bepinat/CAMEL}} \citep{Epinat12} modified to work with iFTS data cubes in wavenumber rather than in wavelength. Both H$\alpha$ and \NII{}$\lambda$6584 lines were simultaneously fitted (\NII{}$\lambda$6548 not covered).
The velocity map is shown in Fig.~\ref{fig:sitelle}.
We also compared the \bino{} velocities with the {\em SITELLE} velocities for 12 overlapping regions, all close to the galaxy. Velocities are always consistent within the combined uncertainties of both measurements.

\section{Results}

\subsection{The velocity field}
\label{subsec:velo}

\ga{}'s western side is the near side and its southern side has lower radial velocity than the northern side \citep[e.g.,][]{Vollmer04,Chemin06}, as expected since its spiral arms are trailing arms. This galactic rotation pattern, as shown in Fig.~\ref{fig:sitelle}, is also observed in stripped clumps (Fig.~\ref{fig:f1}), at least within $\sim$ 50 kpc from the galaxy.
Along the tail, there is a clear radial velocity gradient, starting from the system velocity of the galaxy to roughly the system velocity of the Virgo cluster. This shows clear evidence of stripped clouds being decelerated by the local RPS and, eventually, kinematically mixed with the surrounding ICM, and losing the memory of infall.
The average radial velocity gradient over the projected length of the tail is $\sim$ 5.6~\kms~kpc$^{-1}$, which is about half of that observed for the orphan cloud in A1367 but at smaller scales \citep{Ge21}.

Stripped clouds are subject to both local drag force and the gravity from \ga{} so it is difficult to solve its deceleration history analytically. If we ignore \ga{}'s gravity, the deceleration of the cloud can be written as $d \upsilon / d t = - b \upsilon^{2}$,
where $\upsilon$ is the velocity of the cloud relative to the surrounding medium and $b$ is the deceleration parameter. $b$ can be written as $\rho_{\rm ICM} / (\rho_{\rm cloud} h)$, where $h$ is the length of the cloud along the ram pressure direction and $\rho_{\rm cloud}$ and $\rho_{\rm ICM}$ are the density of the cloud and the ICM respectively. $\rho_{\rm cloud} h$ is also the column density of the cloud. Such an equation can be solved to obtain the relation between the traveled distance and velocity, $l = \ln(\upsilon_{0}/\upsilon)/b$, where $\upsilon_{0}$ is the initial velocity.
The distance of the cloud to the galaxy is then the distance traveled by the galaxy after time $t$ minus the above distance traveled by the stripped cloud.

The above solution is a 1D solution. We need to connect the radial velocity and projected distance on the plane of sky, which can be done by assuming a constant angle with the line of sight for \ga{}'s motion ($\theta$=0 corresponds to motion fully along the line of sight). Thus, the projected distance and the radial velocity of stripped clouds can be written as: $l_{\rm sky} = (\xi - 1 - \ln\xi) \sin\theta / b$, where $\xi = \upsilon_{\rm 0, rad} / \upsilon_{\rm rad}$ and the subscripts sky and rad mean the sky and radial component respectively.
Such a simple model can well describe the observed cloud velocities in the latter half of the tail (e.g., beyond 120 kpc), with $\sin\theta/b =$ 100 - 140 kpc and assuming $\upsilon_{\rm 0, rad}$ = 1300 \kms{}. However the cloud deceleration is too fast close to the galaxy, which is not surprising for the lack of gravity to slow down deceleration in the simple model. 

We can also estimate the timescale when the velocity is reduced to $\sim$ 200 \kms{} from the cluster velocity when the typical ICM bulk motion and turbulence begin to take over \citep[e.g.,][]{2025ApJ...993L..11X}.
The timescale is $\tau = 5.5\sin\theta/(b \upsilon_{\rm 0, rad})$.
For $\sin\theta/b =$ 120 kpc, the timescale is 0.50 Gyr.
This timescale is 2.9 times larger than the time required for a clump to travel 230 kpc with a 1300 km/s velocity (if \ga{}'s projected velocity is similar to its radial velocity), as stripped clouds would follow \ga{} and take time to slow down.
We note that this timescale is longer than the derived timescales of 100 - 300 Myr for the peak of ram pressure stripping from \cite{Vollmer04} and \cite{Boselli06}. However, the H$\alpha$ clumps that are far away from \ga{} may trace the ISM or circumgalactic medium (CGM) stripped long before the peak of ram pressure stripping.
In another VESTIGE paper, Sarpa et al. (2025) will present a detailed kinematic code to study the infall of cluster galaxies like \ga{} more accurately.

We emphasize that the current slit sampling is not unbiased, limited by the slit positioning and the available H$\alpha$ imaging data. Some regions beyond 80 kpc are poorly covered. The true extent of H$\alpha$ clumps may also be longer than what is known now, which requires more observations near the end of the H$\alpha$ tail. There is spread or scatter of velocities at fixed distance from the galaxy, which should come from the combination of remaining galactic rotation and developing of turbulence in the tail.

Because of the incomplete coverage and intrinsic scatter of velocities from galactic rotation and the ICM turbulence, a rigorous modeling is difficult. There is also likely another contamination from Galactic clouds, which could explain some of low-velocity clouds (e.g., those six discussed earlier).
The simple model also ignores the gravity between the stripped clouds and the parent galaxy, which slows down deceleration to make the actual age longer.
The stripped clouds may also get ablated with time, which effective reduces the cloud column density and makes the actual age shorter.
The stripped ISM and CGM can also be mixed with the surrounding ICM \citep[e.g.,][]{Tonnesen21}.

\begin{figure}[t]
\vspace{-0.5cm}
\hspace{-0.35cm}
\includegraphics[width=0.53\textwidth]{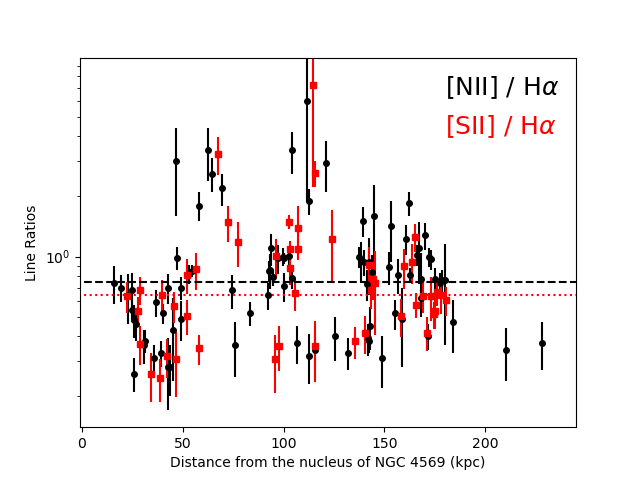}
\vspace{-0.6cm}
\caption{The \NII{}$\lambda$6584\AA{} / \Ha{} ratios for 88 clumps and the \SII{}$\lambda\lambda$6716, 6731 / \Ha{} ratios for 49 clumps trailing \ga{} vs. the distance from the nucleus of \ga{}. The dashed lines show the median ratios. The shown uncertainty is 1-$\sigma$.
}
    \label{fig:line_ratio}
\end{figure}

\subsection{The typical properties of the H$\alpha$ clumps/filaments}

The H$\alpha$ tail is delineated by many clumps/filaments, while a diffuse component with a high covering fraction is absent, at least from the current data. 
The clumps/filaments are often in structure complexes, such as three of them highlighted in Fig.~\ref{fig:f1}.
The complex in zoom-in a consists of clumps with typical size of $\sim$ 0.5 kpc, with fainter emission connecting clumps. Its full extent, pointing back to \ga{}, is $\sim$ 31 kpc. The typical H$\alpha$+\NII{} surface brightness is $\sim$ 5.5$\times10^{-18}$ erg s$^{-1}$ cm$^{-2}$ arcsec$^{-2}$ at $\sim$ 0.6 kpc linear scales. 
The complex in zoom-in b has some long filaments with length up to 8 kpc. While \gb{} is projected nearby, their velocities are very different.
The complex in the right side of zoom-in c has a linear extent of $\sim$ 22 kpc, with a similar surface brightness to other complexes.
We note that the VESTIGE H$\alpha$ data are not sensitive to faint, diffuse emission because of the scattered light and flat field. On the other hand, such a component is absent even close to \ga{} where \Ha{} emission mainly comes from bright filaments, unlike many other RPS galaxies, e.g., ESO~137-001 \citep{Luo23}, many in Coma and A1367 \citep[e.g.,][]{Yagi10,Pedrini22} and some GASP galaxies \citep[e.g.,][]{Tomivic21}.
No matter whether a very faint, diffuse \Ha{} component exists in \ga{}'s tail,
the warm, ionized gas is unlikely the dominant contributor to the mass budget for its small filling factor \citep[e.g.,][]{Sun10,Jachym17,Sun22} so we do not estimate its total mass which also depends on the assumed geometry.

Similar to the ionized gas within 80 kpc from \ga{}, discussed in \cite{Boselli16}, there are no compact young star clusters detected from the Next Generation Virgo Cluster Survey (NGVS) NGVS optical data and the deep {\em Galex} data as discussed in \cite{Boselli16}.
We also attempted such a search with the VESTIGE H$\alpha$ imaging data. We used SEXTRACTOR with the same criteria to select compact emission-line regions as did in \cite{Waldron23}. An H$\alpha$ luminosity lower limit of 5$\times10^{36}$ erg s$^{-1}$ within a circular aperture with a radius of 1.5$''$ is also set, as the existing VESTIGE studies of \hii{} regions have the H$\alpha$ luminosity completeness cut-off at $10^{37}$ erg s$^{-1}$ \citep[e.g.,][]{2025A&A...696A..78B}. \NII{} $\lambda$ 6584 / H$\alpha$ = 0.5 was also assumed. No such emission-line regions are detected beyond 50 kpc from \ga{}'s nucleus in the tail, with the highest H$\alpha$ luminosity at $\sim 3\times10^{36}$ erg s$^{-1}$ in several long filaments.
Thus, SF is unlikely the main ionization mechanism for the tail H$\alpha$ clouds.
We examined the \NII{}/H$\alpha$ and \SII{}/H$\alpha$ ratios when possible, as shown in Fig.~\ref{fig:line_ratio}.
While the Galactic extinction has been corrected, the intrinsic extinction is not corrected, as the ionization mechanism is unclear. The effect on these diffuse regions should be very small for these two ratios.
The \NII{}/H$\alpha$ ratios of these clouds are 0.25 - 3.0 with a median of 0.75. The \SII{}/H$\alpha$ ratios are 0.2 - 3.0 with a median of 0.64. These elevated ratios suggested ionization mechanisms other than young stars, e.g., shocks or mixing with the hot ICM \citep[e.g.,][]{Pedrini22}. 
Low \NII{}/H$\alpha$ and \SII{}/H$\alpha$ ratios also do not necessarily correspond to star forming regions \citep[e.g.,][]{Ge21,Pedrini22}.
Detailed studies of line diagnostics will be included in a future paper.

\section{Discussion and Conclusions}

The dominant role of RPS on the evolution of \ga{} is previously known \citep[e.g.,][]{Vollmer04,Boselli06,Boselli16}.
This work extends the 80 kpc H$\alpha$ tail discovered by \cite{Boselli16} to at least 230 kpc projected distance from the galaxy, while also establishing a coherent radial velocity gradient along the H$\alpha$ clumps/filaments. While NGC~4531 is projected near the end of the long H$\alpha$ tail, it is clearly separated from the nearby H$\alpha$ clumps in velocity space. We also examined the VESTIGE and NGVS deep optical continuum imaging data. No tidal features are detected close to NGC~4531. Thus, there is no evidence for any impact from NGC~4531 on \ga{}'s long H$\alpha$ tail.
We also examined other nearby Virgo cluster galaxies. The H$\alpha$ clump at the end of the tail is 0.79 deg from \ga{} and 0.88 deg from M87.
Within 1 deg of that clump, there are no other Virgo late-type or irregular galaxies that are at least brighter than \gb{}'s magnitude + 1 mag at the $r$ band, and there is no evidence of \Ha{} connection to any galaxies other than \ga{}.

With this discovery, \ga{} now possesses the longest H$\alpha$ RPS tail known.
Before this discovery, the longest RPS tail is UGC~6697 in A1367, with its mostly straight H$\alpha$ tail traced to nearly 150 kpc from the nucleus \citep{Yagi17}. 
CGCG~97-079 in A1367 has its H$\alpha$ tail detected to nearly 140 kpc from its nucleus \citep{Yagi17}. 
A few RPS tails are traced to $\sim$ 90 - 100 kpc from the nucleus, including ESO~137-001 in H$\alpha$ and X-rays \citep{Sun22,Luo23}, NGC~4848 and IC~4040 in H$\alpha$ \citep{Yagi10}, JO~206 in H$\alpha$ \citep{Ramatsoku19} and IC~2276 in radio \citep{Roberts24}.
Two very long and mostly linear stellar tails have also been revealed, Kite
\citep[$\sim$ 380 kpc,][]{Zaritsky23} and GMP~2640
\citep[$\sim$ 250 kpc,][]{Grishin21}, but there is little diffuse gas known and it is unclear whether RPS is the only mechanism there.
The 250 kpc \hi{} tail at the back of FGC~1287 in A1367 may have complicated origin, as the available RP seems to be not strong enough to remove the observed amount of \hi{} gas from the galaxy at its current location \citep{Scott22}.

Our model in Section~\ref{subsec:velo} gives the deceleration timescale of a stripped cloud, which is essentially the same as the drag time for a cold cloud accelerating by a wind, typically found in the studies of ISM and CGM \citep[e.g.,][]{Klein94,Gronke18}. On the other hand, the destruction timescale of the cloud would be $\sim$ ($\rho_{\rm cloud} / \rho_{\rm ICM})^{1/2}$ smaller than the drag time or the deceleration timescale \citep[e.g.,][]{Klein94}. While ($\rho_{\rm cloud} / \rho_{\rm ICM})^{1/2}$ is uncertain, it should be larger than 10, from e.g., the thermal pressure equilibrium argument. Thus, the stripped clouds are not expected to survive for such a long time, or found so far away from the galaxy. This is the classical cloud crushing problem. The typical solution involves mixing with the ICM and the enhanced cooling in the mixing layer \citep[e.g.,][]{Gronke18,Ji19,Tonnesen21}. Thermal conduction may also help to extend the cloud's lifetime \citep[e.g.,][]{Sander21}.
Alternatively, the ISM is stripped as a whole entity and only a small fraction of dense clouds are left behind, as suggested by \cite{Vollmer12}.

While wide-field broad-band optical imaging surveys have become routines, the discovery space for sensitive narrow-band imaging surveys is still vast \citep[e.g.,][]{Lokhorst22,Drechsler23}. The ISM removed from cluster galaxies pollute the hot ICM and the growing number of examples (see the discussion in \citealt{Boselli22} and this example) suggest that those cold/warm ISM can survive for at least Gyr.
\ga{} showcases the potential to use wide-field H$\alpha$ surveys + follow-up optical spectroscopy to study the ICM kinematics and the stripping history of cluster galaxies. Stripped clouds can survive for at least $\sim$ 0.5 Gyr and continue to shine in H$\alpha$. If detected individually far from the parent galaxy (e.g., clouds near the end of \ga{}'s tail), they may resemble the orphan cloud in A1367 \citep{Yagi17,Ge21}. More examples will hopefully be revealed in the future.
Evolution of the stripped clouds in galaxy clusters also allows detailed studies of the related micro transport processes in mixing between the cold, stripped gas and the hot ICM.
Intracluster space may light up in H$\alpha$ once we reach below 10$^{-19}$ erg s$^{-1}$ cm$^{-2}$ arcsec$^{-2}$ at kpc$^2$ scales with the data from e.g., MOTHRA \citep{Lokhorst22}, while the data presented in this work only give a glimpse of the future!

\begin{acknowledgements}
MS thanks Ben Weiner for help on the {\em MMT} observations and data analysis. MS thanks Greg Bryan for helpful discussion.
Support for this work was provided by the NSF grant 2407821. WF acknowledges support from the Smithsonian Institution. This research is also based on observations obtained with the {\em APO} 3.5-meter telescope, which is owned and operated by the Astrophysical Research Consortium. We are grateful to the whole CFHT team who assisted us in the preparation and in the execution of the observations and in the calibration and data reduction: 
Todd Burdullis, Daniel Devost, Bill Mahoney, Nadine Manset, Andreea Petric, Simon Prunet, Kanoa Withington. 
Based on observations obtained with SITELLE, a joint project between Universite Laval, ABB-Bomem, Universite de Montreal, and the Canada-France-Hawaii Telescope (CFHT) with funding support from the Canada Foundation for Innovation (CFI), the
National Sciences and Engineering Research Council of Canada (NSERC), Fond de Recheche du Quebec - Nature et Technologies (FRQNT) and CFHT.
The Canada-France-Hawaii Telescope is operated from the summit of Maunakea by the National Research Council of Canada, the Institut National des Sciences de l'Univers of the Centre National de la Recherche Scientifique of France, and the
University of Hawaii. The observations at the Canada-France-Hawaii Telescope were performed with care and respect from the summit of Maunakea which is a significant cultural and historic site.
This research has made use of the NASA/IPAC Extragalactic Database (NED) which is operated by the Jet Propulsion Laboratory, California Institute of Technology, under contract with the National Aeronautics and Space Administration.
\end{acknowledgements}

\bibliographystyle{aa}
\bibliography{ming}

\end{document}